\documentclass[%
preprint,
secnumarabic,
amssymb, amsmath,
aip,cha,
groupedaddress,
frontmatterverbose,
]{revtex4-1}
\usepackage{graphicx, setspace}
\usepackage[list=true]{subcaption}
\usepackage{docs}%
\usepackage{bm}%
\usepackage[colorlinks=true,linkcolor=blue]{hyperref}
\expandafter\ifx\csname package@font\endcsname\relax\else
\expandafter\expandafter
\expandafter\usepackage
\expandafter\expandafter
\expandafter{\csname package@font\endcsname}%
\fi
\begin{document}
\title{Analytical Estimate of Phase Mixing Time of Longitudinal Akhiezer - Polovin Wave}%
\author{Arghya Mukherjee}
\email{arghya@ipr.res.in}
\author{Sudip Sengupta}
\affiliation{Institute for Plasma Research, Bhat, Gandhinagar 382 428, India}%
%
%
\begin{abstract}
Phase mixing and eventual breaking of longitudinal Akhiezer - Polovin wave subjected to a small amplitude longitudinal perturbation is studied analytically.
It is well known that longitudinal Akhiezer -  Polovin wave
breaks via the process of phase mixing at an amplitude well below its breaking amplitude, when subjected 
to arbitrarily small longitudinal perturbation[Phys. Rev. Lett.108, 125005 (2012)]. 
Here we analytically show that the phase mixing time (breaking time) scales with $\beta$(phase velocity) and $u_m$(maximum fluid velocity) as 
$ \omega_{p}\tau_{mix} \sim \frac{2\pi\beta}{3\delta}[1/u_m^2 - 1/4]$, where $\delta$ is the amplitude of perturbation. This analytical dependence of phase mixing time on $\beta$, $u_m$ and $\delta$ is further verified using numerical simulations based on Dawson sheet model.
\end{abstract}

\maketitle

\section{INTRODUCTION}
Study of space - time evolution of  nonlinear oscillations and waves in cold plasmas and their breaking has been an interesting and fundamental topic of research over several decades.
  Beginning with the pioneering work of Akhiezer and Polovin\cite{akh} and Dawson\cite{daw1} this subject has
 retained its importance till date
 not only due to its fundamental academic interest to nonlinear 
 plasma theory, but also because it serves as a useful paradigm to elucidate the underlying physics behind a wide range of physical phenomena, ranging from laboratory based laser/beam plasma interaction experiments\cite{taj,modena,malka,hegelich,schw,matlis,faure,rech,esa} to
 some astrophysical phenomena\cite{botha,voit}, where large amplitude oscillations and waves are usually encountered.
  
 In 1956, Akhiezer and Polovin\cite{akh} obtained an exact solution representing purely one-dimensional longitudinal
travelling wave in a cold plasma including relativistic mass effects. This travelling wave solution was derived by solving the cold plasma relativistic fluid-Maxwell equations in a stationary wave frame. The maximum amplitude of this wave was shown to be limited by the wave breaking limit which is given by $\frac{eE_{WB}}{m\omega_pc} = \sqrt{2}{(\gamma_{ph} -1)}^{1/2}$. Here $\gamma_{ph} = 
1/\sqrt{1-\frac{v_{ph}^2}{c^2}}$ is the Lorentz factor associated with the phase velocity $v_{ph}$ of the Akhiezer - Polovin wave.
In 1989, Infeld and Rowlands\cite{inf} presented an exact space - time dependent solution for the relativistic cold plasma fluid- Maxwell equations in Lagrange coordinate. The solution presented by them shows explosive behaviour for all initial 
conditions except for the one which are needed to excite Akhiezer - Polovin waves.
Physically, this explosive behaviour arises due to the relativistic electron mass variation which causes the characteristic electron plasma frequency to acquire 
spatial dependencies, due to which neighbouring electrons get gradually out of phase and eventually cross causing the wave to break\cite{guptapre,guptappcf,chandan1,aipp}.
This process of wave breaking through gradual phase mixing is not exhibited by a pure Akhiezer - Polovin wave.

Recently Verma \textit{et al.}\cite{verma1} constructed longitudinal Akhiezer - Polovin travelling wave solutions from the exact space time
dependent solution of Infeld and Rowlands using appropriate choice of initial conditions. 
It was further shown by the same authors\cite{verma2} that even a longitudinal Akhiezer - Polovin wave 
breaks through the gradual process of phase mixing at an amplitude
well below its wave breaking limit, when it is subjected to arbritrarily small longitudinal perturbations. 
It was found through extensive numerical simulations that phase mixing time decreases with increasing $u_m$( for a fixed $\delta$ ) 
and increasing $\delta$( for a fixed $u_m$ ), where $u_m$ and $\delta$ are respectively the velocity amplitudes of the Akhiezer - Polovin wave and the applied 
perturbation.
In this paper we present a weakly relativistic calculation which analytically brings out the scaling of  phase mixing time with the parameters of the Akhiezer - Polovin wave 
( $u_m$ and $\beta$; $\beta$ is the phase velocity of the Akhezer - Polovin wave ) and the amplitude of the perturbation $\delta$.
 We have further verified our scaling numerically using a code
 based on Dawson sheet model\cite{daw1,dawpof}, 
 which shows a good agreement between numerical and analytical 
 results.
 
 In section II, we briefly describe the Dawson sheet model and construct longitudinal Akhiezer -Polovin travelling wave solution using this model\cite{aipp}. 
 This derivation although equivalent to that presented in ref. {\cite{verma1}, \cite{verma2}} is physically more transparent. 
 It is presented here for the sake of completeness. Section III is devoted to weakly relativistic calculations required for the estimation 
 of phase mixing time of longitudinal Akhiezer - Polovin wave subjected to small amplitude sinusoidal perturbation. 
 In section IV analytical predictions are compared with numerical findings. Finally section V contains discussions and 
 summary of our results.
 
\section{RELATIVISTIC TRAVELLING WAVE SOLUTION }
According to the Dawson sheet model description of a cold plasma, electrons are assumed to be infinite sheets of charges embedded
in a cold immobile positive ion background\citep{dawpof}. Evolution of any coherent mode can be studied in terms of oscillating motion of these sheets about their equilibrium positions.
Let $x_{eq}$ and $\xi(x_{eq},\tau)$ respectively be the equilibrium position and displacement from the equillibrium position
of an electron sheet. In terms of $x_{eq}$ and $\xi(x_{eq},\tau)$ the associated fluid quantities \textit{viz} number density, velocity and electric field can respectively be written as $n(x_{eq},\tau) = n_0/(1+\partial \xi(x_{eq},\tau)/\partial x_{eq})$, $v(x_{eq},\tau) = \dot \xi$ and $E(x_{eq},\tau) = 4\pi en_0\xi(x_{eq},\tau)$. Here dot represents differentiation w.r.t Lagrange time $\tau$. These expressions can further be represented in terms of Euler coordinates ($x$,$t$) using the transformations $x = x_{eq} + \xi(x_{eq},\tau)$ and $t = \tau$. Thus, for any given initial condition, once $\xi(x_{eq},\tau)$ for a particular sheet is computed as a function of $x_{eq}$ and $\tau$, the problem of evolution of a coherent mode in space and time is essentially solved in principle. This can be obtained by solving the relativistic equation of motion of a sheet which using Gauss's law, may be written as
\begin{equation}
\frac{\ddot \xi}{{(1-\dot \xi^2)}^{3/2}} + \xi = 0    \label{eq:1}
\end{equation}
where $\tau \rightarrow \omega_p\tau$, $\xi \rightarrow \frac{\omega_p\xi}{c}$, $\dot \xi \rightarrow \frac{\dot \xi}{c}$,
$E \rightarrow \frac{eE}{m\omega_pc}$, $\omega_p$ is the nonrelativistic plasma frequency.
 Multiplying Eq.\ref{eq:1} by $\dot \xi$
, we get
\begin{equation}
\frac{1}{{(1-\dot \xi^2)}^{1/2}} + \frac{\xi^2}{2} = a(x_{eq}) \label{eq:2}
\end{equation}
Here \textquotedblleft$\emph a$\textquotedblright corresponds to the total energy of the sheet. Substituting
 \begin{equation}
\xi = \sqrt{2(a-1)}sin\alpha \label{eq:3}
\end{equation}
solution of Eq.\ref{eq:1} becomes
 \begin{equation}
\tau = \frac{2}{r^{\prime}}E(\alpha,r) - r^{\prime} F(\alpha,r) + \Phi(x_{eq}) \label{eq:4}
\end{equation}
which gives $\alpha$ as an implicit function of $\tau$ and $x_{eq}$, where $r = {[(a-1)/(a+1)]}^{1/2}$ and 
$r^{\prime} = \sqrt{1 - r^2}$. Eq.\ref{eq:3} along with Eq.\ref{eq:4} describes the motion of an electron sheet about its equilibrium position for a given set of initial conditions $\Phi(x_{eq})$ and $r(x_{eq})$. The frequency $\omega$ of an electron sheet is obtained by integrating Eq.\ref{eq:2} between two turning points ($\dot \xi = 0)$ and is given by 
\begin{equation}
\omega = \frac{\pi}{2}\frac{r^{\prime}}{[2E(r) - r^{\prime 2}K(r)]} \label{eq:5}
\end{equation}
It is evident from Eq.\ref{eq:5} that for arbitrary set of initial conditions, \textquotedblleft$\omega$\textquotedblright 
is in general a function of \textquotedblleft$\emph x_{eq}$\textquotedblright. Since any coherent mode is made up of a large number of electron sheets oscillating about their equilibrium positions, this spatial 
dependency of $\omega$ causes the neighbouring electron sheets to gradually go out of phase with time, which eventually leads to crossing of electron sheet trajectories resulting in singularities in the electron density profile. This is the phenomenon of phase mixing leading to wave breaking. For a sinusoidal initial density profile and for wave like initial conditions, the phenomenon of phase mixing leading to wave breaking is convincingly demonstrated in references\cite{guptapre,guptappcf,inf,drake}. 

As stated in the introduction, in ref.\cite{verma1} it is shown, that it is possible to choose a special set of initial conditions which excites a propagating solution with phase velocity $\beta$, which does not phase mix and break. This propagating solution is nothing but a longitudinal Akhiezer-Polovin wave.  
Absence of phase mixing implies, from Eq.\ref{eq:5}, that
\textquotedblleft$\emph a$\textquotedblright (energy of an oscillating sheet) should be independent of \textquotedblleft$\emph x_{eq}$\textquotedblright and propagation with a fixed phase velocity $\beta$ fixes the functional form of $\Phi(x_{eq})$ as $\Phi(x_{eq}) = x_{eq}/\beta$ .
This form of $\Phi(x_{eq})$ is obtained by choosing $\xi$ ( hence $\alpha$ ) to be entirely a function of $\psi$ = $\omega(t - x/\beta)$. 
Thus the initial conditions for exciting a longitudinal Akhiezer-Polovin wave are
\begin{equation}
\xi_{ap}(x_{eq},0) = \frac{2r}{r^\prime}sin\alpha  \label{eq:6}
\end{equation}
\begin{equation}
\dot \xi_{ap}(x_{eq},0) = \frac{2r}{{r^\prime}^2}\frac{cos\alpha\sqrt{1-r^2sin^2\alpha}}{1+\frac{2r^2}{{r^\prime}^2}cos^2\alpha}  \label{eq:7}
\end{equation}
along with $\alpha(x_{eq},0)$ implicitly given by
\begin{equation}
\frac{2}{r^{\prime}}E(\alpha,r) - r^{\prime} F(\alpha,r) = -\frac{x_{eq}}{\beta} \label{eq:8}
\end{equation}
Following Akhiezer - Polovin's work\cite{akh} we now choose $u_m$(maximum fluid velocity) and $\beta$ as 
independent parameters, instead of $\textit{a}$(or \textit{r}) and $\beta$. $\textit{a}$ and $u_m$ are related to each other through Eq.\ref{eq:1} and Eq.\ref{eq:2} as 
$a = 1/\sqrt{1-{u_m}^2}$.
In the next section we add a small perturbation to $\xi_{ap}(x_{eq},0)$ and $\dot \xi_{ap}(x_{eq},0)$ 
which leads to phase mixing and subsequent breaking of longitudinal Akhiezer - Polovin wave.
\section{ESTIMATION OF PHASE MIXING TIME}
Adding a small amplitude sinusoidal perturbation of amplitude $\delta$ and wavelength 
$k_{ap}$(same
as the longitudinal Akhiezer - Polovin wave) to $\xi_{ap}$ and $\dot \xi_{ap}$, we get
\begin{equation}
\xi_{per} = \xi_{ap} -  \frac{\delta}{\omega_{ap}}sin(k_{ap}x_{eq}) \label{eq:9}
\end{equation}
\begin{equation}
\dot \xi_{per} = \dot\xi_{ap} +\delta cos(k_{ap}x_{eq}) \label{eq:10}
\end{equation}
where $\xi_{ap}$ and $\dot \xi_{ap}$ are the required initial condition for exciting a longitudinal AP wave\cite{verma1} and $\xi_{per}$, $\dot \xi_{per}$ are the perturbed initial conditions.
The perturbed initial conditions are equivalent to adding a small amplitude sinusoidal density perturbation propagating with phase velocity $\beta$, to longitudinal Akhiezer - Polovin wave.
In the weakly relativistic limit, keeping terms linear in $\delta$, the energy associated with an electron sheet becomes
\begin{equation}
a_{per} \approx a+\left[-\frac{\xi_{ap}\delta}{\omega_{ap}}sin(k_{ap}x_{eq}) + \delta \dot\xi_{ap} cos(k_{ap}x_{eq})\right] \label{eq:11}
\end{equation}
where $a \approx 1 + \xi^2/2 + \dot \xi^2/2 $. 
This in turn gives
\begin{eqnarray}
r_{per}^2 &\approx & r^2-\frac{2r^2}{a^2-1}\delta\left[\frac{\xi_{ap}}{\omega_{ap}}sin(k_{ap}x_{eq}) 
- \dot\xi_{ap} cos(k_{ap}x_{eq}) \right] \label{eq:12}
\end{eqnarray}
Finally substituting $r^2_{per}$ in Eq.\ref{eq:5}, the frequency of oscillation in the weakly relativistic limit stands as
\begin{eqnarray}
\omega_{per} &\approx & 1-\frac{3r^2}{4}
+\frac{3r^2}{2(a^2-1)}\delta\left[\frac{\xi_{ap}}{\omega_{ap}}sin(k_{ap}x_{eq}) 
-\dot\xi_{ap} cos(k_{ap}x_{eq})\right] \label{eq:13}
\end{eqnarray}
It is clear from the above expression that the frequency of the wave is dependent on the equilibrium position 
of the electrons which leads to the phenomena of phase mixing\cite{verma2}.
Following Dawsons' argument\cite{daw1}, the phase mixing time ($\tau_{mix}$) depends on the spatial derivative of frequency as $\tau_{mix} \approx \pi/2\xi_{max}(d\omega_{per}/dx_{eq})$.
Differentiating Eq.\ref{eq:13} w.r.t $x_{eq}$ and using $\xi_{max} = 2r/r^\prime$, the phase mixing time in this case becomes

\begin{equation}
\tau_{mix} \approx \frac{2\pi \beta}{3\delta}\left[\frac{1}{u_m^2} - \frac{1}{4}\right] \label{eq:14}
\end{equation}
It is clear from the above expression that phase mixing time scales directly with $\beta$, inversely with $\delta$ and 
has a $u_m^{-2}$ dependence on $u_m$.

In the next section we verify these predictions numerically using Dawson sheet simulation.
\section{NUMERICAL RESULTS}
Using a code based on Dawson sheet model, we numerically verify the process of phase mixing of a large amplitude
Akhiezer - Polovin wave perturbed by a small amplitude sinusoidal perturbation. We first load Akhiezer - Polovin type initial condition with a sinusoidal 
perturbation of amplitude $\delta$ in a one-dimensional relativistic sheet code containing $\sim$ 10000 
electron sheets. Using these initial conditions the equation of motion for each sheet is then solved using
fourth order Runge-Kutta scheme. At each time step, ordering of the sheets is checked for sheet crossing(electron trajectory crossing). Phase mixing
time is measured as the time taken by any two of the adjacent sheets to cross over. We terminate our code at this time
because the expression for electric field ($E = 4\pi en_0\xi$) used in equation of motion(Eq.\ref{eq:1}) becomes invalid beyond this point\citep{daw1,guptaprl}.

\begin{figure}[htbp]
\center
\includegraphics[width=1\textwidth]{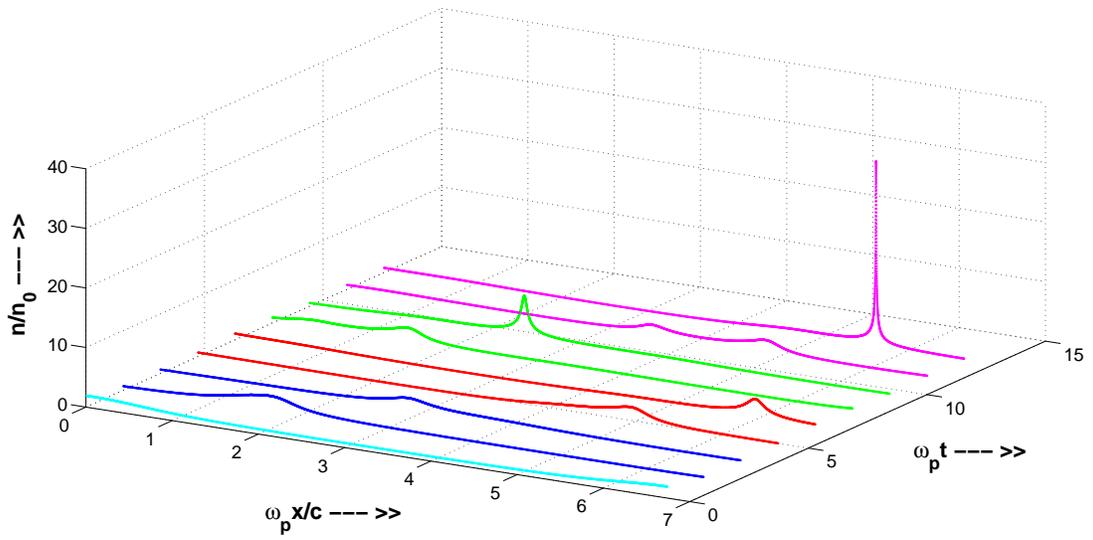}
\center
\caption{(Color online) Space-time evolution of the electron density for an Akhiezer - Polovin wave with velocity amplitude $u_m$ = 0.55
 with perturbation amplitude $\delta$ = 0.1 and $\beta$ = 0.9995}  
\label{fig:fig0} 
\end{figure}

\begin{figure}[htbp]
\includegraphics[width=1\textwidth]{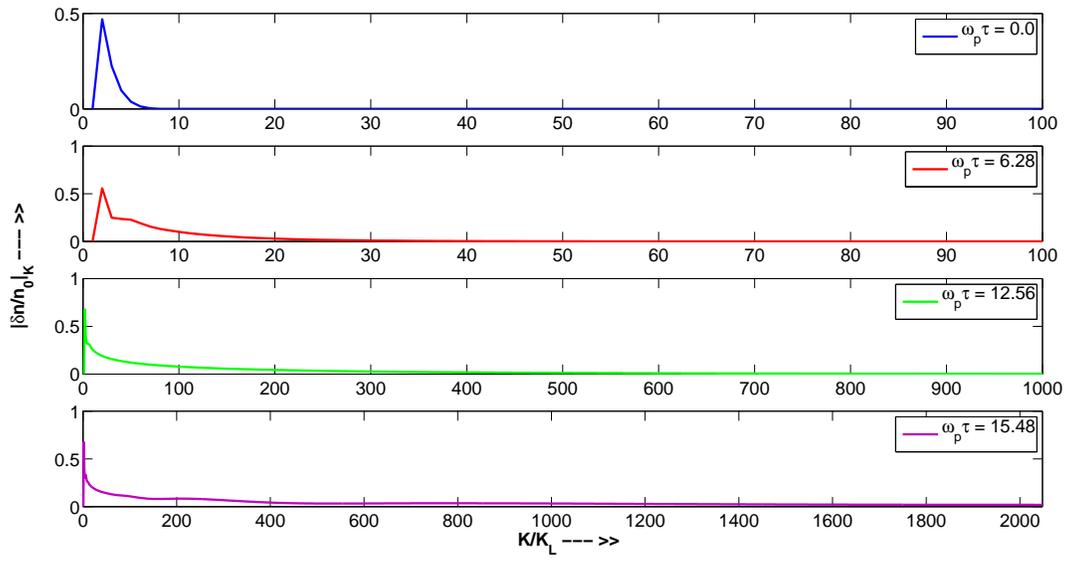}  
\caption{(Color online) Fourier spectrum of a Akhiezer - Polovin wave with velocity amplitude $u_m$ = 0.55 with perturbation amplitude $\delta$ = 0.1 and $\beta$ = 0.9995
at different time steps.}  
\label{fig:fig1} 
\end{figure}

\begin{figure}[htbp] 
\includegraphics[width=1\textwidth]{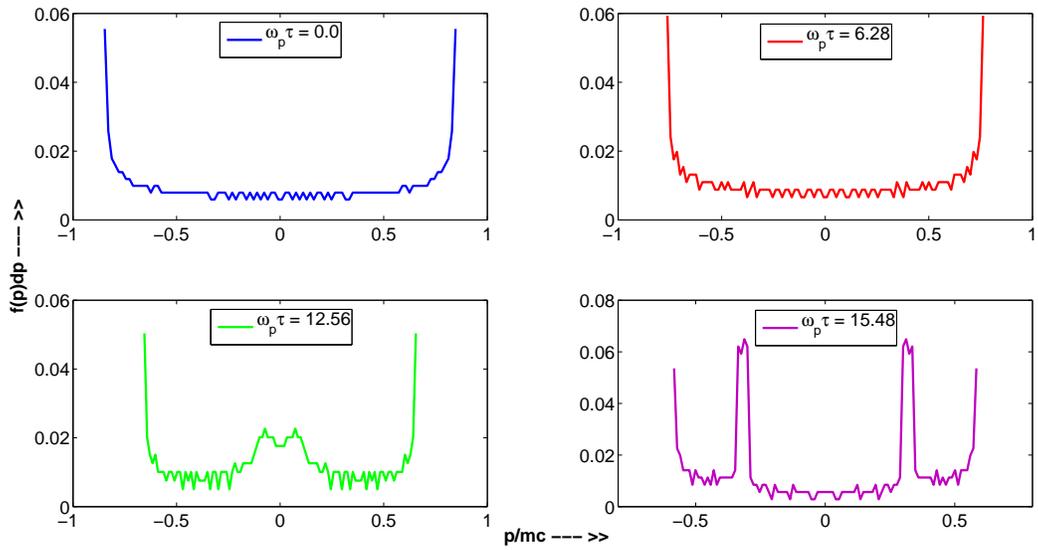}
\caption{(Color online) Momentum distribution of a Akhiezer - Polovin wave with velocity amplitude $u_m$ = 0.55 with perturbation amplitude $\delta$ = 0.1 and $\beta$ = 0.9995
at different time steps.}  
\label{fig:fig2} 
\end{figure}

Fig-\ref{fig:fig0} shows the space time evolution of the electron density
profile of the resultant structure. As time progresses, the density profile becomes more and more spiky as energy is irreversibly
transferred to higher and higher harmonics. A manifestation of this process is also seen in the Fourier spectrum where we see a broadening of the spectrum as time progresses(Fig-\ref{fig:fig1}). Thus the energy which was loaded in the primary mode eventually distributes over higher modes. The interaction of these high \textquotedblleft$\emph k$\textquotedblright   modes with the particles (sheets) accelerates the particles, causing the initial delta-function momentum distribution to spread.
Fig-\ref{fig:fig2} shows that as time progresses, the momentum distribution function gradually spreads generating multi-stream flow; a clear indication of phase mixing leading to breaking. Figure. 4 - 6 respectively show the variation of phase mixing time  with respect to $\delta$, $u_m$ and $\beta$ for fixed values of the other two parameters. In all the cases points represent the simulation results and the solid line represents our scaling obtained from Eq.\ref{eq:14}. In all cases, the analytical expression(Eq.\ref{eq:14}) shows a very good fit to the observed numerical results, thus vindicating our weakly relativistic calculation.

\section{DISCUSSIONS AND SUMMARY}
 The phenomenon of phase mixing is a manifestation of spatially-dependent plasma frequency\citep{guptappcf}.
 It is well known that large amplitude longitudinal AP wave
breaks via the process of phase mixing at an amplitude well below the breaking amplitude for AP wave$(\sqrt 2 
\sqrt{\gamma_{ph} - 1})$
, when subjected 
to arbritrarily small longitudinal perturbation\cite{verma2}. We have derived an expression for
this phase mixing time which brings out its dependence on $u_m$, $\beta$ and $\delta$. Our weakly relativistic calculation indicates that the phase mixing time scales linearly with $\beta$, inversely with $\delta$ and has $1/u_m^2$ dependence on $u_m$. We have verified our scaling using numerical simulations.

We note here that, the dependence of phase mixing time on $\Delta$ (density amplitude, 
$(\delta n/n_0)_{max}$ can be obtained from Eq.\ref{eq:14} as 
${\tau_{mix} \approx \frac{2\pi \beta}{3\delta}\left[{\left(\frac{1+\Delta n}{\beta \Delta n}\right)}^2 - \frac{1}{4}\right]}$ 
by eliminating $u_m$ using $n = n_0\beta/(\beta - u)$. This shows that for $\Delta \gg 1$, $\tau_{mix}$ is essentially independent of $\Delta$ and for $\Delta \ll 1$, $\tau_{mix}$ scales as
$\sim 1/\delta\Delta^2$. Fig-\ref{fig:subcaption7},\ref{fig:subcaption8} show the dependence of phase mixing time on $\Delta$. We emphasize here that for $\Delta \ll 1$, and $\delta \sim \Delta$, our expression for phase mixing time exhibits $\sim 1/\Delta^3$ scaling, in conforming with the results preented in references\citep{guptapre,inf,gor}.

\newpage
\listoffigures

\newpage
\vspace*{\fill}
\vspace*{\fill}

\newpage
\vspace*{\fill}
\vspace*{\fill}

\newpage
\vspace*{\fill}
\vspace*{\fill}

\newpage
\vspace*{\fill}
\begin{figure}[htbp]
\begin{subfigure}{.45\linewidth}
\centering
\includegraphics[width=1\textwidth]{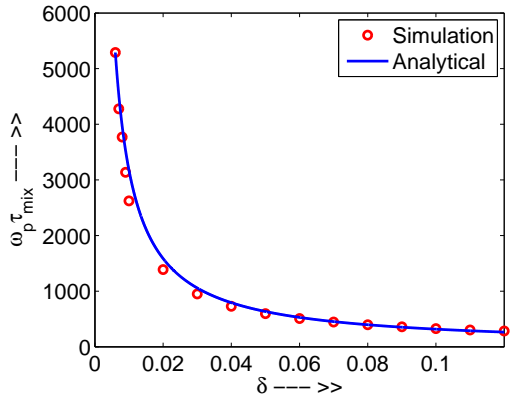}
\caption{}
\label{fig:subcaption1}
\end{subfigure}%
\quad
\begin{subfigure}{.45\linewidth}
\centering
\includegraphics[width=1\textwidth]{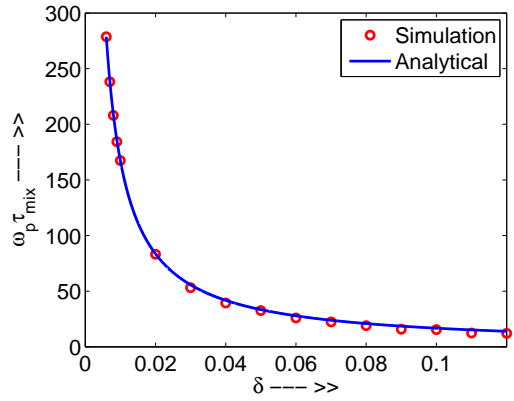}
\caption{}
\label{fig:subcaption2}
\end{subfigure}
\caption{(Color online) Analytical (solid) and numerical (circles) scalings of the phase mixing time for a finite amplitude Akhiezer - Polovin wave for 
$u_m$ = 0.20(\ref{fig:subcaption1}), 0.55(\ref{fig:subcaption2}) and $\beta$ = 0.9995 as a function of perturbation amplitudes $(\delta)$} 
\end{figure}
\vspace*{\fill}

\newpage
\vspace*{\fill}
\begin{figure}[htbp]
\begin{subfigure}[b]{.45\linewidth}
\centering
\includegraphics[width=1\textwidth]{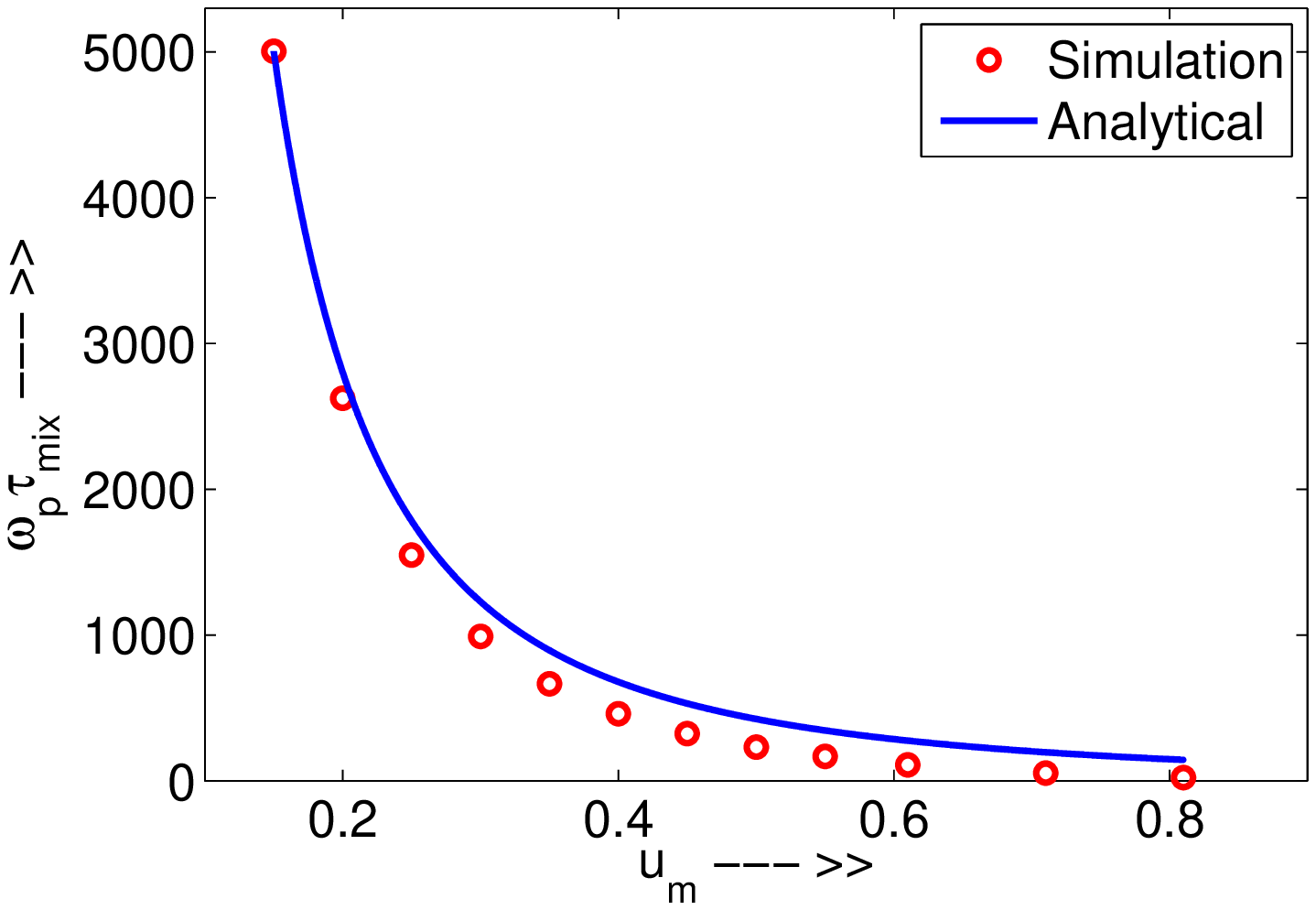}
\caption{}
\label{fig:subcaption3}
\end{subfigure}%
\quad
\begin{subfigure}[b]{.45\linewidth}
\centering
\includegraphics[width=1\textwidth]{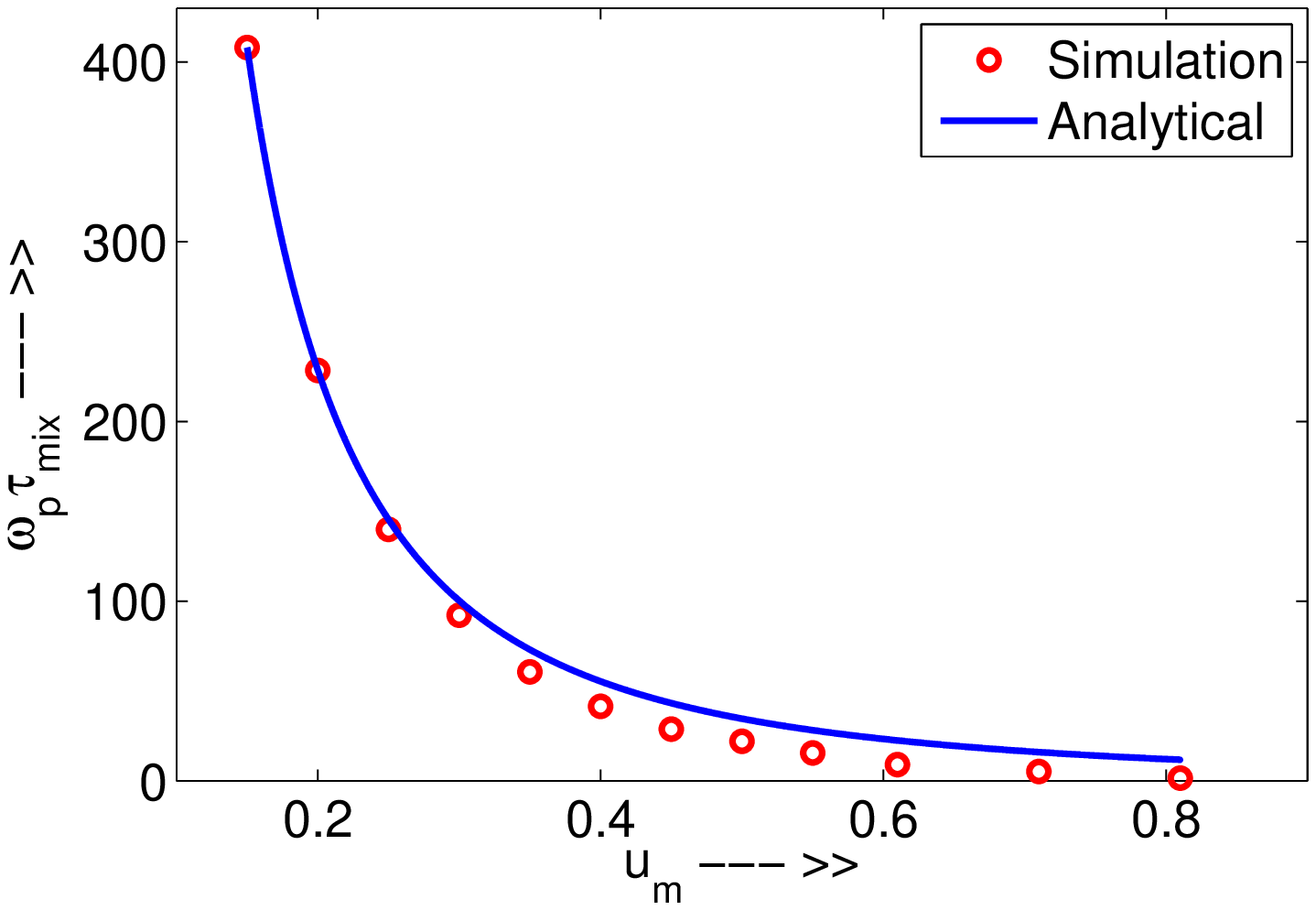}
\caption{}
\label{fig:subcaption4}
\end{subfigure}
\caption{(Color online) Analytical (solid) and numerical (circles) scalings of the phase mixing time as a function of the
amplitude of Akhiezer - Polovin wave ($u_m$) in the presence of a finite perturbation$(\delta)$ = 0.01(\ref{fig:subcaption3}),0.1(\ref{fig:subcaption4}) and $\beta$ = 0.9995.}
\end{figure}
\vspace*{\fill}

\newpage
\vspace*{\fill}
\begin{figure}[htbp]
\begin{subfigure}[b]{.45\linewidth}
\centering
\includegraphics[width=1\textwidth]{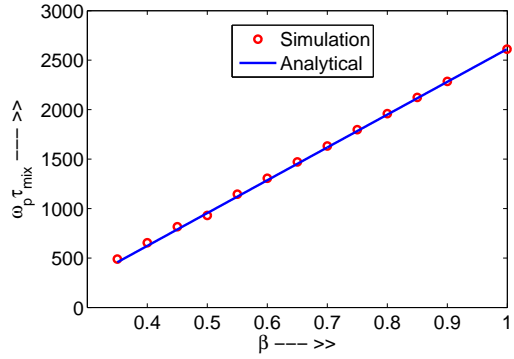}
\caption{}
\label{fig:subcaption5}
\end{subfigure}%
\quad
\begin{subfigure}[b]{.45\linewidth}
\centering
\includegraphics[width=1\textwidth]{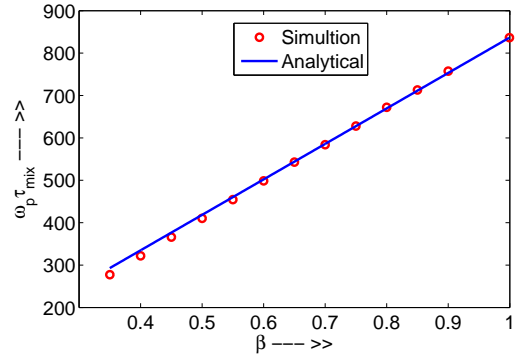}
\caption{}
\label{fig:subcaption6}
\end{subfigure}
\caption{(Color online) Analytical (solid) and numerical (circles) scalings of the phase mixing time as a function of the phase
velocity $\beta$ for a fixed amplitude of Akhiezer - Polovin wave $u_m$ = 0.2, $\delta$ = 0.01(\ref{fig:subcaption5}) and $u_m$ = 0.1, $\delta$ = 0.1(\ref{fig:subcaption6})}.
\end{figure}
\vspace*{\fill}

\newpage
\vspace*{\fill}
\begin{figure}[htbp]
\begin{subfigure}[b]{.45\linewidth}
\centering
\includegraphics[width=1\textwidth]{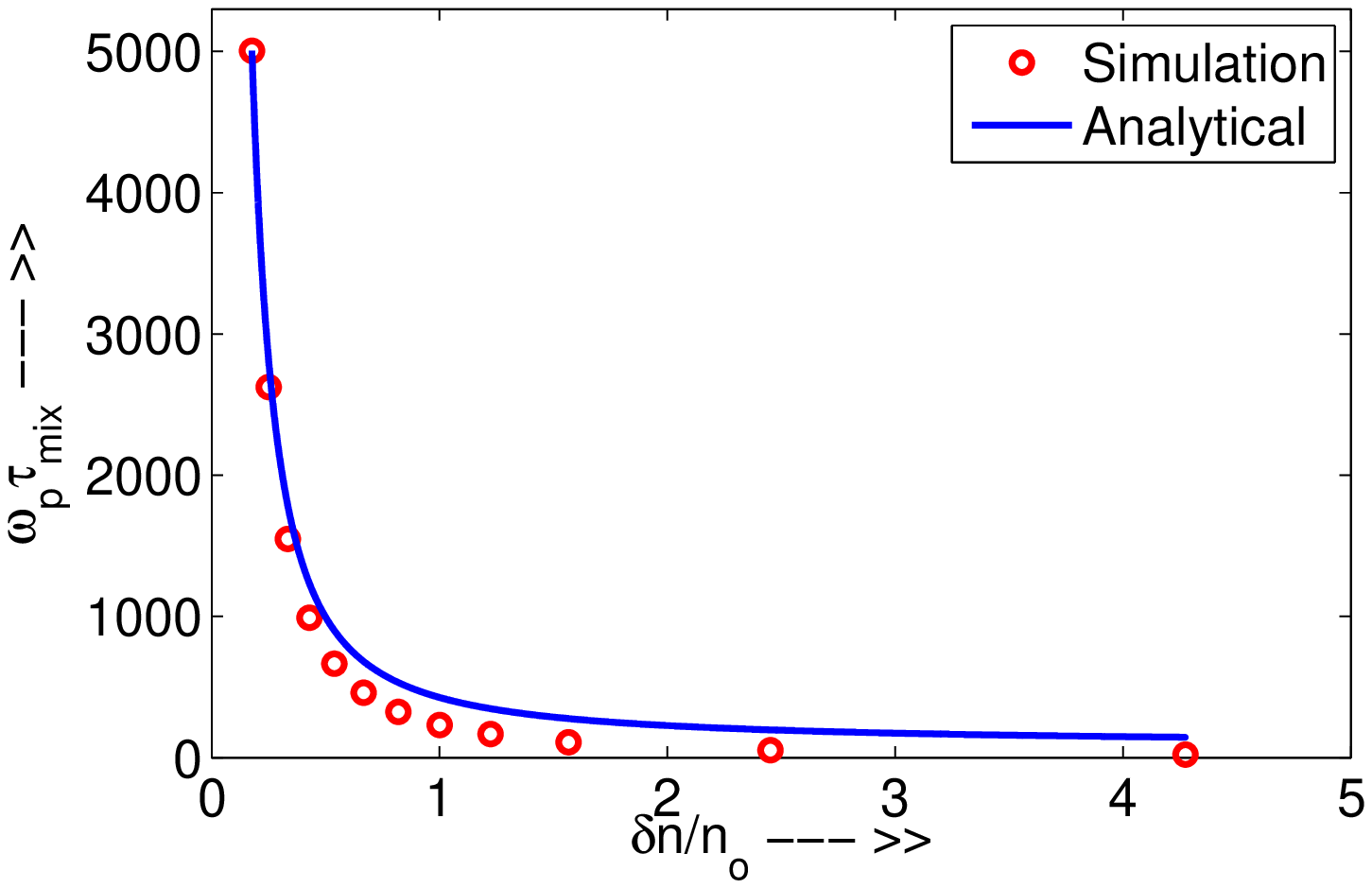}
\caption{}
\label{fig:subcaption7}
\end{subfigure}%
\quad
\begin{subfigure}[b]{.45\linewidth}
\centering
\includegraphics[width=1\textwidth]{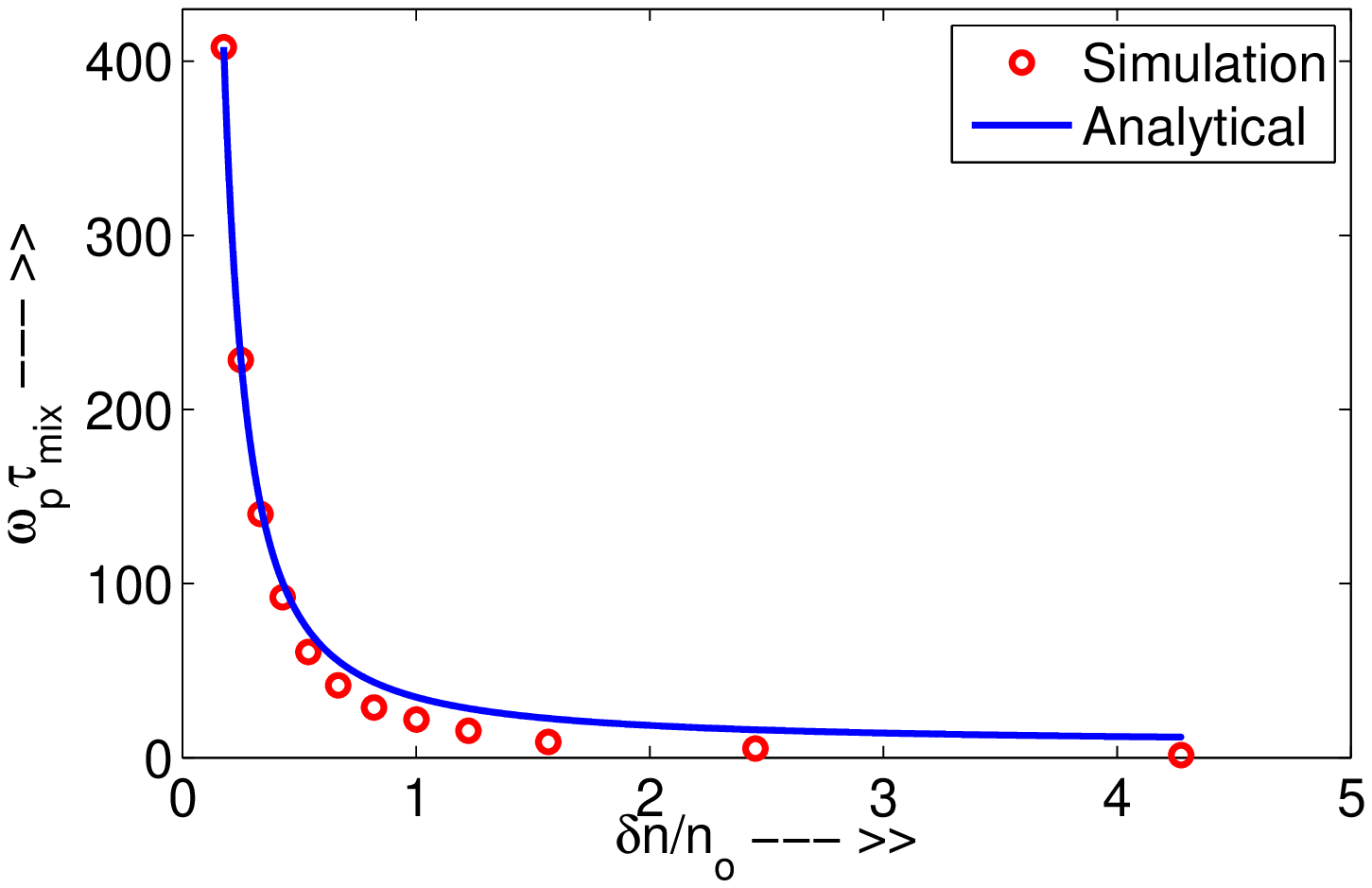}
\caption{}
\label{fig:subcaption8}
\end{subfigure}
\caption{(Color online) Analytical (solid) and numerical (circles) scalings of the phase mixing time as a function of the density
amplitude of the Akhiezer - Polovin wave $(\delta n/n \sim \Delta n)$ in the presence of a finite perturbation$(\delta)$ = 0.01(\ref{fig:subcaption7}),0.1(\ref{fig:subcaption8}) and $\beta$ = 0.9995}
\end{figure}
\vspace*{\fill}
\end{document}